\documentclass[twocolumn,showpacs,preprintnumbers,revsymb,prb, superscriptaddress]{revtex4}

\usepackage[dvips]{graphicx}
\usepackage{bm}
\usepackage{mathptmx}

\begin{document}

%\title{Comment on ``Dynamical corrections to the DFT-LDA
%           electron conductance in nanoscale systems''}
\title{Treatment of electron viscosity in quantum conductance}
\author{J. Jung}
\affiliation{Physics Division, National Center for Theoretical Sciences,  
	P.O. Box 2-131, Hsinchu, Taiwan}
\author{P. Bokes } 
\affiliation{Department of Physics, Faculty of Electrical Engineering and
        Information Technology, Slovak University of Technology, 
	Ilkovi\v{c}ova 3, 812 19 Bratislava, Slovak Republic}
\affiliation{Department of Physics, University of York, Heslington, York
         YO10 5DD, United Kingdom}
\author{R. W. Godby}
\affiliation{Department of Physics, University of York, Heslington, York
         YO10 5DD, United Kingdom}
\date{\today{}}

\maketitle

In a recent paper Sai {\it et al.} [1] identified a 
correction $R^{dyn}$ to the DC conductance of nanoscale 
junctions arising from dynamical exchange-correlation ($XC$) effects 
within time-dependent density functional theory. This
quantity contributes to the total resistance through
$R=R_{s}+R^{dyn}$
where $R_{s}$ is the resistance evaluated in the absence of 
dynamical $XC$ effects. 
In this Comment we show that the numerical estimation of 
$R^{dyn}$ in example systems of the type they considered 
should be considerably reduced, once a more appropriate form for the 
shear electron viscosity $\eta$ is used. 

Sai {\it et al.}'s expression for $R^{dyn}$, based on electron-liquid theory [2], is a
one-dimensional integral between the two electrodes
\begin{equation}
R^{dyn}=\frac{4}{3e^{2}A_{c}}\int_{a}^{b}\eta \frac{\left( \partial
_{z}n\right) ^{2}}{n^{4}}dz 
\end{equation}
where $A_{c}$ is the cross-sectional area, $\eta $ is the shear viscosity of the
electron liquid, and $n$ is the electron density. 
The example system we have considered is the metal-vacuum-metal (MVM)
junction that can be formed by two jellium surfaces separated by a distance $d$ [3].
Since this system is translationally invariant parallel to the surface, 
the electron density $n(z)$ is defined unambiguously. We have chosen
$r_S=3$ for the jellium electrodes, to allow 
comparison with the gold-electrode systems presented in Ref. 1.
The density is calculated within the LDA.

Sai {\it et al.} used a {\it constant} shear viscosity $\eta_c =%
\hbar  ( k_{F} / \pi a_{0})^{3/2} / 120$
corresponding to formula (4.7) of Ref. 2, appropriate only in the 
high-density weakly inhomogeneous limit, and based on the bulk electrode density. 
Here the quantities $%
k_{F}=(3\pi ^{2}n)^{1/3}$ and $a_{0}$ are the Fermi wavevector of the bulk
electrodes and the Bohr radius respectively.
We make two changes to this.  
First, we use Formula (4.10) of Ref. 2,
$\eta_v \simeq n / (
60r_{S}^{-3/2}+80r_{S}^{-1}-40r_{S}^{-2/3}+62r_{S}^{-1/3})$,
more appropriate for realistic densities ($r_s=0...20$ a.u.), [2]
in recognition of the fact that $r_S=3$ is not a high density.
This alone
reduces the dynamical resistance by a factor of 5.37 for a density corresponding to $r_S=3$.
Second, in Eq. (1) we evaluate this viscosity at the {\it local} density rather than taking the
bulk viscosity to apply outside the bulk region.  This 
further reduces the dynamical resistance by a $d$-dependent factor of 1.16--18.4 (see Table I), 
particularly for larger $d$ when the dominant contribution to the integral in Eq. (1) 
comes from the low-density region in the vacuum (see Fig. 1).
\begin{table}
\begin{center}
\begin{tabular}{rccc}
$d$ & $R_{s}$ & $R_{c}^{dyn}$ & $R_{v}^{dyn}$     \\ 
\hline
1   &   117   &  1.71  &  0.275      \\ 
3   &   241   &  48.5  &  5.01       \\
5   &   712   &  530   &  29.4       \\
9   &   14900 &  59600 &  604      
\end{tabular}
\caption{Non-interacting and dynamical resistances per unit area
  (a.u.) as a function of the separation $d$ (a.u.).   The interpolated local formulation of the viscosity, 
  $\eta_v$, considerably reduces the dynamical resistance ($R_{v}^{dyn}$) relative to the high-density
  bulk formulation of Ref. 1 ($R_{c}^{dyn}$).} 
\end{center}
\end{table}
\begin{figure}
\includegraphics[width=10cm,angle=0]{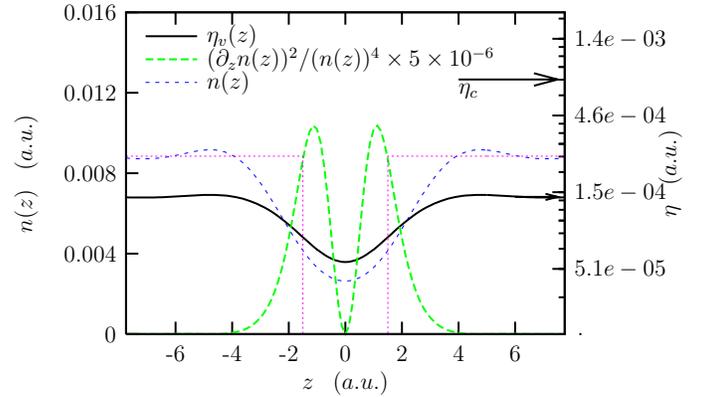}
\caption{(Color-online) 
MVM junction separated at $d=3$ a.u. 
The electron charge density and the background positive charge density are
represented by thin discontinuous lines. 
Important contributions to the integral of
Eq. (1) (green thick dashed line) arise within the vacuum junction, 
where the electronic viscosity (solid line) is lower,
tending to reduce the dynamical resistance. 
The value of viscosity used in Ref. 1 is indicated by $\eta_c$.}
\end{figure}
Thus a more appropriate choice of the shear electron viscosity $\eta$ reduces the dynamical
resistance by a factor between 6 and 98, causing it to become very small compared with
the single-particle resistance in all cases studied.

%\begin{acknowledgments}
This work was partially funded by the EU {\it Nanoquanta}
NoE (NMP4-CT-2004-500198). P. Bokes acknowledges support from the Slovak grant
agency VEGA (project No. 1/2020/05) and the NATO Security through Science
Programme (EAP.RIG.981521).
%\end{acknowledgments}

\end{document}